# Purcell effect in Hyperbolic Metamaterial Resonators

Alexey P. Slobozhanyuk[1,3*], Pavel Ginzburg[2,7**], David A. Powell[1], Ivan Iorsh[3], Alexander S. Shalin[3,4,5],

Paulina Segovia[2], Alexey V. Krasavin[2], Gregory A. Wurtz[2], Viktor A. Podolskiy[6], Pavel A. Belov[3] and

Anatoly V. Zayats[2]

[1]Nonlinear Physics Center, Research School of Physics and Engineering, Australian National University, Canberra ACT 0200, Australia

[2]Department of Physics, King's College London, Strand, London WC2R 2LS, United Kingdom

[3]ITMO University, St. Petersburg 197101, Russia

[4]Kotel'nikov Institute of Radio Engineering and Electronics of RAS (Ulyanovsk branch), Ulyanovsk 432011, Russia

[5]Ulyanovsk State University, Ulyanovsk 432017, Russia

[6]Department of Physics and Applied Physics, University of Massachusetts Lowell, One University Ave, Lowell, MA, 01854, USA

[7]Department of Physical Electronics, Fleischman Faculty of Engineering, Tel Aviv University, Tel Aviv 69978, Israel

The radiation dynamics of optical emitters can be manipulated by properly designed material structures providing high local density of photonic states, a phenomenon often referred to as the Purcell effect. Plasmonic nanorod metamaterials with hyperbolic dispersion of electromagnetic modes are believed to deliver a significant Purcell enhancement with both broadband and non-resonant nature. Here, we have investigated finite-size cavities formed by nanorod metamaterials and shown that the main mechanism of the Purcell effect in these hyperbolic resonators originates from the cavity hyperbolic modes, which in a microscopic description stem from the interacting cylindrical surface plasmon modes of the finite number of nanorods forming the cavity. It is found that emitters polarized perpendicular to the nanorods exhibit strong decay rate enhancement, which is predominantly influenced by the rod length. We demonstrate that this enhancement originates from Fabry–Pérot modes of the metamaterial cavity. The Purcell factors, delivered by those cavity modes, reach several hundred, which is 4-5 times larger than those emerging at the epsilon near zero transition frequencies. The effect of enhancement is less pronounced for dipoles, polarized along the rods. Furthermore, it was shown that the Purcell factor delivered by Fabry–Pérot modes follows the dimension parameters of the array, while the decay rate in the epsilon near-zero regime is almost insensitive to geometry. The presented analysis shows a possibility to engineer emitter properties in the structured metamaterials, addressing their microscopic structure.

Corresponding authors: *a.slobozhanyuk@phoi.ifmo.ru, **pginzburg@post.tau.ac.il

## 1. Introduction

The local density of optical states (LDOS) of photonic modes can strongly affect quantum dynamics of light-matter interactions [1]. Macroscopic electromagnetic structures lead to reshaping of free space electromagnetic modes and, as a result, can lead to either a local enhancement or reduction of the interaction strength. The rate of spontaneous emission in a weak light-matter coupling regime, calculated on the basis of the Fermi golden rule, is proportional to the LDOS and its change relative to free space is referred to the name of Purcell factor [2]. Furthermore, the formalism of the Purcell effect can be generalized to higher-order effects, such as spontaneous two-photon emission [3,4]. The Purcell enhancement in dielectric cavities is typically related to the ratio of the quality factor of the resonance to the volume occupied by the resonant mode. Various types of photonic cavities can deliver quality factors as high as $10^{10}$ and satisfy the conditions to reach the strong coupling regime [5] where the Purcell factor description of decay dynamics breaks down [6]. Noble metal (plasmonic) nanostructures provide relatively low quality factors but yield sub-wavelength optical confinement [7,8] and as a result, also efficiently influence spontaneous emission [9,10]. This nanoplasmonic approach is extremely beneficial for certain quantum optical applications, where improved and designed scattering cross-sections are required to develop functionalities at the nanoscale and single photon levels [11]. The Purcell enhancement in plasmonic nanostructures is resonant, and thus has a limited bandwidth [12]. Moreover, it depends significantly on the relative position of the emitters with respect to a metal nanostructure, posing serious challenges and limitations for practical implementations.

A qualitatively different approach for Purcell effect engineering relies on designing the hyperbolic dispersion in anisotropic metamaterials, which ensures high nonresonant Purcell factors in a broad wavelength range [13, 14]. These metamaterials with extreme anisotropy of dielectric permittivity, also known as hyperbolic metamaterials, have recently attracted significant attention due to their unusual electromagnetic properties. Homogenised hyperbolic metamaterials were theoretically shown to provide infinitely large LDOS and, as the result, are expected to deliver extremely high Purcell enhancements [15,16]. This diverging LDOS originates from the hyperbolic dispersion of modes in uniaxial crystals, having opposite sign of the permittivity components in the ordinary and extraordinary directions, perpendicular and parallel to the optical axis, respectively. The fundamental limitations for this type of enhancement result from a particular metamaterial realisation as composites of finite-length scale components, commonly referred to as "meta-atoms"[17], as well as the metamaterials' nonlocal response [18,19]. The most widely used realizations of hyperbolic metamaterials are based on layered metal–dielectric structures [20] or vertically aligned nanorod arrays [21]. Hyperbolic metamaterials also served as building blocks for optical components with enhanced capabilities, such as hyperbolic cavities [22] and waveguides [23].

In this work we analyse the emission properties of a radiating dipole embedded inside or in a close proximity of a finite-size three-dimensional resonator formed by a nanorod-based hyperbolic metamaterial. Taking into account the details of the hyperbolic metamaterial realization as a finite number of plasmonic nanorods, we show that the Purcell enhancement originates from Fabry–Pérot modes of the resonator formed by a hyperbolic metamaterial. The role of the modes of the hyperbolic resonators on the Purcell factor was investigated for different resonator sizes, and the importance of the emitter's position within the resonator has been considered. We also demonstrate that for fixed rod diameter and inter-rod separation, the Purcell enhancement in a 16x16 nanorod array converges to that of the infinite metamaterial slab (infinite number of finite length rods). This enables comparing the Purcell enhancement provided by both finite-sized and infinite structures and separating the impact of the modal structure of finite-size resonators.

## 2. Effective medium theory and numerical modeling

We consider the metamaterial consisting of a square array of plasmonic (Au) nanorods (Fig. 1). This basic configuration enables addressing all the relevant effects, with substrate material and embedding dielectric material straightforwardly included in numerical modeling. In the first approximation, neglecting nonlocal effects [24], the optical response of such a structure can be obtained from a homogenization procedure of the nanorod composite [25], representing it as an effective uniaxial medium with permittivity tensor

$$\varepsilon = \begin{pmatrix} \varepsilon_{xx} & 0 & 0 \\ 0 & \varepsilon_{xx} & 0 \\ 0 & 0 & \varepsilon_{zz} \end{pmatrix},$$

where $\varepsilon_{xx} = \varepsilon_{yy}$ and $\varepsilon_{zz}$ are the permittivities for the light polarization perpendicular to and along the nanorod axes, respectively. In the frequency range where $\varepsilon_{xx}$ and $\varepsilon_{zz}$ have opposite signs (Fig. 1(c)), extraordinary electromagnetic modes propagating in such an anisotropic medium have hyperbolic dispersion. The frequency range where the real part of the effective permittivity $\varepsilon_{zz}$ becomes vanishingly small is called the epsilon near-zero (ENZ) regime. For the considered system, this crossing occurs at around 520 nm wavelength (Fig. 1(c)).

The numerical simulations have been performed using the time domain solver of the CST Microwave Studio 2014 package [26]. We used perfect matched layers (PML) boundary and additional space was added between the structure and the PMLs in order to prevent evanescent waves from interacting with the boundaries. The subwavelength dipolar emitter was modeled as a perfect electric conductor (PEC) nanorod of length 28 nm and radius 1 nm. The Purcell factor was calculated through an input impedance of a point dipole source. As was previously shown [27, 28], this calculation method is completely equivalent to the Green's function approach, which is widely used in photonics [29].

## 3. Results

### 3.1. Hyperbolic metamaterial resonators.

First, the modal structure of finite-size resonators made of homogenized hyperbolic metamaterial based on the nanorod assembly will be derived analytically. The Purcell factor is proportional to the imaginary part of the Green's function in a medium. In the case of the *x*-oriented electric dipole (Fig. 1), the Purcell factor is given by

$$R(\omega) = \mathrm{Im}\, G(\mathbf{r}, \mathbf{r}, \omega) \approx \mathrm{Im} \sum_{n,l,m,\sigma} \frac{|E_x^{n,l,m,\sigma}(\mathbf{r})|^2}{\omega^2 - \omega_{n,l,m,\sigma}^2} \qquad , \tag{0.0}$$

where (*n,l,m*) are integers denoting the eigenmode number within the cavity, made of the metamaterial, and $\sigma = \{\mathrm{TE}, \mathrm{TM}\}$ is the mode's polarization, where TE corresponds to the modes with the electric field lying in the *xy* plane and TM corresponds to the modes with the magnetic field in the *xy* plane.

The rigorous eigenmode analysis of the anisotropic and lossy rectangular resonator requires sophisticated numerical techniques. However, approximate expressions for the eigen frequencies and field distributions in the resonator can be derived within an approximate analytical formalism [30]. Within this formalism, perfect electric conductor (PEC) boundary conditions are imposed at the side walls of the resonator at $x = \pm L_x / 2, y = \pm L_y / 2$ in order to obtain the resonator modes numbers, corresponding to quantized $k_{x,(m)}$ and $k_{y,(l)}$ wavevectors in *x* and *y* directions, respectively:

$$\begin{aligned} k_{x,(m)} &= \frac{\pi m}{L_x}, \ m = 1,3,5... \\ k_{y,(l)} &= \frac{\pi l}{L_y}, \ l = 1,3,5... \end{aligned} \tag{1.2}$$

If the radiating dipole is situated at the (0,0,z) axis, only the modes with symmetric $E_x$ and $E_y$ field distributions with respect to inversions $x \to -x, y \to -y$ can be excited by the dipole and will contribute to the Purcell effect.

In order to obtain quantization in the remaining z-direction, a slab waveguide is then considered confined in z-direction which may support TM and TE guided modes with the propagation constant $k_{\wedge,(m,l)}$ which should satisfy condition

$$k_{\wedge,(m,l)} = \sqrt{k_{x,(m)}^2 + k_{y,(l)}^2} \,. \tag{1.3}$$

This $k_{\wedge,(m,l)}$ propagation constant can be evaluated by finding the modes of a hyperbolic-metamaterial-slab waveguide in the EMT approximation [31]. As has been shown in the analysis of the metamaterial waveguides, and can also be seen from the numerical modeling below, this approximation holds for lower-order highly confined modes of sufficiently large resonators.

First, we will consider quasi-TE modes of the resonator. By substituting Eq. (1.2) into the dispersion equation for a slab waveguide which is symmetric with the respect to the $z \rightarrow -z$ inversion, we find two classes of modes, which either have the tangential electric field symmetric ( $n = 0, 2, 4, ...$ ) or antisymmetric ( $n = 1, 3, 5...$ ) along the z axis, respectively:

$$\frac{\sqrt{k_{\perp,(m,l)}^2 - \left(\frac{\omega_{m,l,n,\mathrm{TE}}}{c}\right)^2}}{\sqrt{\varepsilon_{xx}\left(\frac{\omega_{m,l,n,\mathrm{TE}}}{c}\right)^2 - k_{\perp,(m,l)}^2}} = \tan\left(\frac{\sqrt{\varepsilon_{xx}\left(\frac{\omega_{m,l,n,\mathrm{TE}}}{c}\right)^2 - k_{\perp,(m,l)}^2}\, L_z}{2} + \frac{\pi}{4}\left(1 - (-1)^n\right)\right). \tag{0.0}$$

If the dipole is placed at $z = 0$, the antisymmetric modes will have the node of the electric field at the dipole position and thus cannot be excited and will not contribute to the Purcell effect (Eq. 1). When the dipole is shifted from the $z = 0$, both symmetric and antisymmetric modes will contribute to the Purcell factor. Similarly, using the waveguide dispersion for TM polarized modes, we obtained

$$\frac{\sqrt{\varepsilon_{xx}}\sqrt{k_{x,(m)}^2 + k_{y,(l)}^2 - \left(\frac{\omega_{m,l,n,\mathrm{TM}}}{c}\right)^2}}{\sqrt{\left(\frac{\omega_{m,l,n,\mathrm{TM}}}{c}\right)^2 - \frac{k_{\perp,(m,l)}^2}{\varepsilon_{zz}}}} = \tan\left(\frac{\sqrt{\varepsilon_{xx}}\sqrt{\left(\frac{\omega_{m,l,n,\mathrm{TM}}}{c}\right)^2 - \frac{k_{\perp,(m,l)}^2}{\varepsilon_{zz}}}\, L_z}{2} + \frac{\pi}{4}\left(1 - (-1)^n\right)\right), \tag{1.5}$$

In this case, however, the modes are symmetric ($n = 0, 2, 4, ...$) or antisymmetric ($n = 1, 3, 5...$) with respect to the tangential component of the magnetic field, with the electric field having opposite symmetry.

In order to obtain a clear physical picture, the spectrum of the eigenmodes will be first analyzed assuming vanishing Ohmic losses in the metamaterial. In this approximation, following Eq. (1.4) the TE mode eigenfrequencies should occupy the interval

$$\frac{\pi}{L_x}\sqrt{(m)^2 + (l)^2} < \frac{\omega_{m,l,n,\text{TE}}}{c} < \frac{\pi}{L_x}\sqrt{\varepsilon_{xx}}\sqrt{(m)^2 + (l)^2} \, , \qquad (1.6)$$

stemming from the requirement that the left hand side of Eqs. (1.2, 1.4) should be real-valued. Therefore, for any finite frequency range, only a finite number of pairs (*m,l*) exists that satisfy Eq.(1.6). For each (*m,l*) pair, a finite set of mode numbers *n* can be found from the solution of Eq.(1.3). Thus, in a finite frequency range, only a finite number of eigenmodes (*m,l,n*) of the metamaterial resonator exists.

Specifically, for the metamaterial resonator of the square cross-section with $L_x = L_y$ = 900 nm and $L_z$ = 350 nm and the effective permittivity as in Fig. 1 (c), the following eigenmodes can be excited in the spectral range from 500 to 1500 nm: TE$_{110}$ at $\lambda$=1450 nm, TE$_{130}$ and TE$_{310}$ at $\lambda$= 780 nm, and TE$_{330}$ at $\lambda$= 570 nm. It should be noted that while the predicted higher-order modes were observed in the rigorous numerical simulations of the nanorod composite, the fundamental mode in the vicinity of 1450 nm has not been observed and occurs at wavelengths larger than 1500 nm. This is a known discrepency [32] which is related to the fact that the simplified analysis used above works worse for the fundamental modes with lower confinement within a resonator, and thus the actual frequency of the TE$_{111}$ mode frequency can deviate substantially from the value predicted by the simplified analytical formalism.

Contrary to the case of the TE modes, for the eigenfrequencies of TM modes a decrease with the increase of $m$ and $l$ as can be seen from Eq. (1.5). This property of the hyperbolic resonators has been observed both theoretically and experimentally [22], and can be understood from the requirement for the TM eigenfrequencies analogous to Eq. (1.6):

$$\frac{\pi^2}{L_x^2}[m^2 + l^2] > \frac{\omega_{m,l,n,\mathrm{TM}}^2}{c^2} > \frac{\pi^2}{L_x^2 \varepsilon_{zz}}[(m)^2 + l^2]. \tag{1.7}$$

Since $\varepsilon_{zz}$ is negative, the right-hand-side inequality holds for any frequency and $m,l$. Thus, there exist modes with arbitrary large $m,l$ that satisfy the left-hand-side inequality. The number of the supported modes is however limited due to the metamaterial realization as a periodic nanorod array. Contrary to the case of the uniform hyperbolic metamaterial, the $x$ and $y$ wavevectors should be within the first Brillouin zone of the array, $k_{x,y} < \pi/a$, where $a$ is the period of the array. Thus, for TM modes, $m$ and $l$ eigenvalues can be 1, 3, 5, and 7 in the case of the 16x16 nanorod array with the parameters as in Fig. 1. For simplicity, in these analytical calculations we do not consider possible coupling between TE and TM modes due to three-dimensional geometrical confinement (the numerical modelling include all the effects).

For each value of $m$ and $l$ there is a number of eigenmodes corresponding to different $n$. This number is finite and increases with $m,l$. Despite a large number of the TM polarized eigenmodes supported by the resonator, many of them have a minor contribution to the overall Purcell factor, since those modes are either characterized by small Q-factors, due to the large damping inside the resonator when the losses in metal are considered, or by a small value of $x$ component of the electric field at the dipole's position, due to the different symmetry properties of the eigenmodes. Namely, some of the eigenmodes would have nodes of the x component of the electric field at the dipole position and some would have antinodes [33]. Moreover, in the vicinity of the ENZ frequency, the modes with large values of $n$ are excited. However, these modes have large losses and, thus, give little contribution to the overall Purcell effect. It should be noted, however, that calculations of the Purcell enhancement for emitters placed in the contact with lossy media, face several challenges as the Green's functions diverge [34]. This problem is usually addressed by introducing a depolarization volume (a small lossless cavity) around the emitter [34]. The numerical modeling below does not, however, face the above issues, as the emitter is placed in the lossless space between actual rods, forming the metamaterial.

The dependence of the resonant wavelength on the resonator length $L_z$ for three modes, $TE_{130}$, $TM_{151}$, and $TM_{551}$ is shown in Figure (2a). The higher-order mode $TM_{551}$ is lower in frequency than the lower order mode $TM_{151}$ as is expected for the hyperbolic resonators. The dependence of the resonant wavelength on the resonator width $L_x$ at the fixed resonator length $L_z$ =350 nm is shown in Figure (2b). As we can see, the resonant wavelengths of the TM modes decrease with the increase of the resonator lateral size. This behavior is evident from Eq. (1.3), since it can be seen that for the fixed value of $k_z^{TM}$ the resonant frequency should increase with increasing $L_x$. In contrast, the wavelengths of the TE modes increase with the increase of the resonator width $L_x$ similar to the case of a conventional anisotropic dielectric resonator.

### 3.2. Purcell enhancement due to the hyperbolic resonator modes

The analytical analysis, performed above, does not account for either the microscopic structure of the metamaterial or the radiation leakage. We now compare the effective medium analytical description to the results of the numerical modeling in the case of an *x*-polarized dipole placed in the centre of the 16x16 array of Au nanorods (period $a = 60$ nm and radius $r = 15$ nm) which corresponds to the resonator dimensions $L_y = L_x = 900$ nm and $L_z = 350$ nm. We have considered both real losses in gold and low losses (artificially reduced in 10 times) for comparison to the analytical model (Fig. 3). The analytical model provides a clear description of the numerical results and the individual eigenmodes can be resolved. The highest Purcell factor corresponds to the excitation of the $TM_{551}$ eigenmode in the vicinity of 1000 nm. The Purcell factor near the ENZ frequency range is not characterized by extremely large values, as was expected for the case of an infinitely large metamaterial [34]. This is due to the fact that (i) the number of modes is limited by the finite period of the array and (ii) higher-order modes are characterized by larger damping, as discussed above.

### 3.3. Saturation of the Purcell enhancement in finite size arrays

The Purcell factor for the electric dipole placed in the centre of the metamaterial resonator was numerically calculated for different sizes of the resonator (Fig. 4). Both orientation of the emitting dipole, parallel and perpendicularly to the nanorods, were analysed in both square and rectangular resonators with up to 18 rods in one direction. The obtained dependence of the Purcell factor shows a fast convergence with an increasing number of rods in the array. In fact, in the arrays larger than 16x16 rods (900x900 nm), the Purcell factor approaches the values for the infinite (in $x$ and $y$ directions) planar metamaterial slab, so that the Purcell factor for 16x16 and 18x18 arrays is essentially the same without the signatures of the resonant modes of the resonator due to the damping of the modes (Fig. 4(a)). Rectangular nanorod arrays show similarly fast convergence, enabling to state that the behavior of 16x16 nanorod structures is extremely close to that of the infinite metamaterial (Fig. 4(c)). In particular, the $x$-oriented dipole source which is located in the central part of the array can excite only even modes (Fig. 4(a)). It can be seen that for the 2x2 array the highest Purcell factor (around 500) is reached at 950 nm wavelength (due to the small number of the rods forming the resonator, the identification of the mode structure of the resonator is not possible in the effective medium formalism as in Section 3.1). For larger arrays this mode exhibits a slight shift to longer wavelengths as expected from Eq. (1.4), and the Purcell factor decreases up to the value of 200. The Purcell value for a $z$-oriented dipole is very low but also follows the mode structure of the resonator with the number of rods (Fig. 4(b)). The rectangular nanorod arrays provide a Purcell factor of around 200 already for 2 rows of nanorods (Fig. 4(c)). The contribution of different transverse modes of the rectangular array in the Purcell factor is more pronounced for the arrays with smaller number of rows and becomes indistinguishable for the arrays with 6 or more rows of rods (Fig. 4 (c), green curve). For all considered sizes of the resonators, a Purcell factor of less than 100 is observed in the ENZ regime, at around 520 nm wavelength (Fig. 1(c)).

### 3.4. Purcell enhancement dependence on the rods length

Now the impact of the resonator height (rod length) on the Purcell factor (Fig. 5) will be investigated. For all nanorod heights, there is a relatively small peak in the vicinity of the ENZ frequency related to the high modal density of bulk plasmon-polariton modes at this frequency [31,35]. The highest observed Purcell factor strongly depends on the rod length. Its maximum shifts to longer wavelengths with the increase in rod height, in accordance with the frequency shift of the resonator mode (Eq. 1.4). This ($TM_{551}$) mode has a characteristic field distribution inside the resonator (Fig. 5(b)) with three pronounced maxima, typical of the 2nd Fabry-Perot mode along the rods. Away from the modes of the resonator, the electric field has a characteristic cross-shaped form (Fig. 5 (c)) typical for a radiating dipole field distribution in a hyperbolic dispersion regime [14].

### 3.5. Purcell enhancement in small hyperbolic resonators.

If, starting from a single nanorod, the number of rods in the resonator is gradually increased, a nontrivial behaviour of the Purcell factor is observed (Fig. 6). The highest Purcell factor is obtained neither with a single rod nor in the limit of an infinite number of rods. The optimal structure provides a resonant mode with a high LDOS which enhances the decay rate. It can be seen that the dipole positioned near the centre of the single nanorod excites the second-order mode n=2 with three maxima of the electric field (Fig. 6(b)) at the wavelength of around 855 nm and the fourth mode n=4 with five field maxima (Fig. 6(c)) at the wavelength of around 610 nm. By adding more nanorods to the resonator, and, thus, changing its size and the modal structure, the resonant frequencies is slightly shifted to the red, in accordance with Eq. 1.4. In particular, for the geometry with four nanorods, the second mode is excited at a wavelength of ca. 940 nm (Fig. 6(d)) and the fourth at a wavelength of ca. 640 nm (Fig. 6(e)). As mentioned above, the dipole located near the middle section of the nanorod layer can only couple to even modes.

**3.5. Purcell enhancement dependence on the dipole position.**

In order to understand the average Purcell factor for an ensemble of randomly distributed emitters, the position-dependent Purcell factor has been investigated. When the position of the emitter is changed along the nanorod length from just outside the metamaterial towards the centre of the metamaterial layer, the Purcell factor has four maxima, which correspond to the four lowest modes (Fig. 7). The odd modes were not excited by the dipole situated at the central point of the array due to symmetry-induced selection rules. It can be seen from the LDOS spectrum (Fig. 7(a)) that the efficiency of the excitation of the resonator modes depends on the local field strength of a particular mode at the position of the radiating dipole. It can be seen that the dipole position corresponds to $(TM_{130})$, $(TM_{551})$ and $(TM_{552})$ modes with two (Fig. 7(b)), three (Fig. 7(c)) and four (Fig. 7(d)) field maxima, respectively, at the wavelengths of 1363 nm, 1000 nm and 750 nm, respectively. For the shorter wavelength of 600 nm, the electric field is shaped as an inverted "V" (Fig. 7(e)), corresponding to the ENZ regime where many modes converge and a quasi-continuum of states is present. The Purcell factor drops off very quickly with increasing the distance between the dipole and the metamaterial surface: dipoles situated more than 20 nm away from the interface do not exhibit any significant Purcell enhancement (Fig. 7).

**5. Conclusion**

A comprehensive numerical and analytical analysis of the Purcell enhancement in finite-size nanorod metamaterial resonators was performed. Using a nanorod metamaterial with hyperbolic dispersion of electromagnetic modes, the resonators with a complex hierarchy of modes can be realized. We have shown that the modes of the hyperbolic resonator are responsible for the enhancement of spontaneous emission rates of emitters placed inside the resonator. Thus, a controllable Purcell enhancement can be achieved in the desired wavelength range by choosing appropriate resonator sizes. Detailed analysis of various types of geometrical arrangement of the metamaterial and emitter was carried out. The results suggest that finite-size metamaterial resonators with properly designed modes outperform infinite metamaterials in terms of radiation efficiency enhancement. It was shown that the influence of only 16x16 nanorod array on the dipole emission properties converges to that of an infinite metamaterial. Our work can provide guidelines for modeling and optimization of experimental samples. As for an outlook for possible future applications, it is worth mentioning nanostructured light-emitting devices with high-speed switching rates, cavities for surface plasmon amplification by stimulated emission of radiation (SPASERs), and sensing applications.

Acknowledgments:

This work has been supported, in part, by EPSRC (UK), the ERC iPLASMM project (321268), US Army Research Office (W911NF-12-1-0533) and Australian Research Council. APS acknowledges support via the SPIE Scholarship. A. P. S., D. A. P., I. I., A. S. S, P. A. B. acknowledge support Australian Research Council (Australia). P.S.O. is grateful for the support from CONACYT. G.W. acknowledges support from the EC FP7 project 304179 (Marie Curie Actions). A.Z. acknowledges support from the Royal Society and the Wolfson Foundation and rejects any involvement in the grants, listed below. The work of A. P. S., D. A. P., I. I., A. S. S, P. A. B. was supported by the Ministry of Education and Science of the Russian Federation, by the Government of the Russian Federation (Grant 074-U01), by Russian Fund for Basic Research (N13-02-00623) and the Dynasty Foundation (Russia).

Figure Captions

Figure 1. (Colour online) (a) Schematic view of the hyperbolic metamaterial resonator. (b) Schematics of the numerical setup. An emitting dipole is inserted in the centre of the resonator. (c)

The effective permittivity of the metamaterial calculated for an infinite array of nanorods with $L_z$ = 350 nm, $a = 60$ nm, $r = 15$ nm (Au permittivity was taken from [36]).

Figure 2. (Colour online) Dependence of the resonant wavelength of the eigenmodes on (a) $L_z$ for the fixed $L_x = L_y$ = 900 nm and (b) $L_x$ for fixed $L_z$ = 350 nm. Error bars indicate the width of the resonance.

Figure 3. (Colour online) The comparison of the numerical (red and blue lines) and analytical (green and black lines) of the Purcell factor in the case of real losses (solid lines) and reduce losses Im($\varepsilon$)/10 (dashed lines). The metamaterial parameters are as in Fig. 1. The resonator size is 16x16 nanorods $L_x$ = $L_y$ = 900 nm.

Figure 4. (Colour online) The Purcell factor dependence on the number of nanorods in square (a,b) and rectangular (c) lattices for an emitting dipole perpendicular (a,c) and parallel (b) to the nanorods. The dipole is located in the centre of the array.

Figure 5. (Colour online) (a) The Purcell factor dependence on the height of the hyperbolic metamaterial resonator (16x16 nanorod array). (b,c) The electric field ($E_x$) distributions excited by the dipole positioned at the centre of the resonator with $L_z$ = 350 nm, at the wavelength of (b) 1000 nm and (c) 600 nm.

Figure 6. (Colour online) (a) Comparison of the Purcell factor for different numbers of plasmonic rods forming a resonator. Electric fields ($E_x$) of the dipole near a single nanorod (b,c) and near four nanorod array (d,e) at the wavelengths of 856 nm (b), 611 nm (c), 937 nm (d) and 637 nm (e).

Figure 7. (Colour online) (a) The Purcell factor dependence on an emitter position inside the metamaterial resonator with $L_z$ = 350 nm. The coordinate z=0 corresponds to the edge of the hyperbolic medium. (b-e) The electric field ($E_x$) distributions excited by the dipole positioned at the $z = 0$ for the wavelengths of (b) 1363 nm (TM$_{130}$ mode), (c) 1000 nm (TM$_{551}$ mode), (d) 750 nm (TM$_{552}$ mode), and (e) 600 nm (nonresonant wavelength). All other parameters as in Fig. 3.

List of Figures

Figure 1

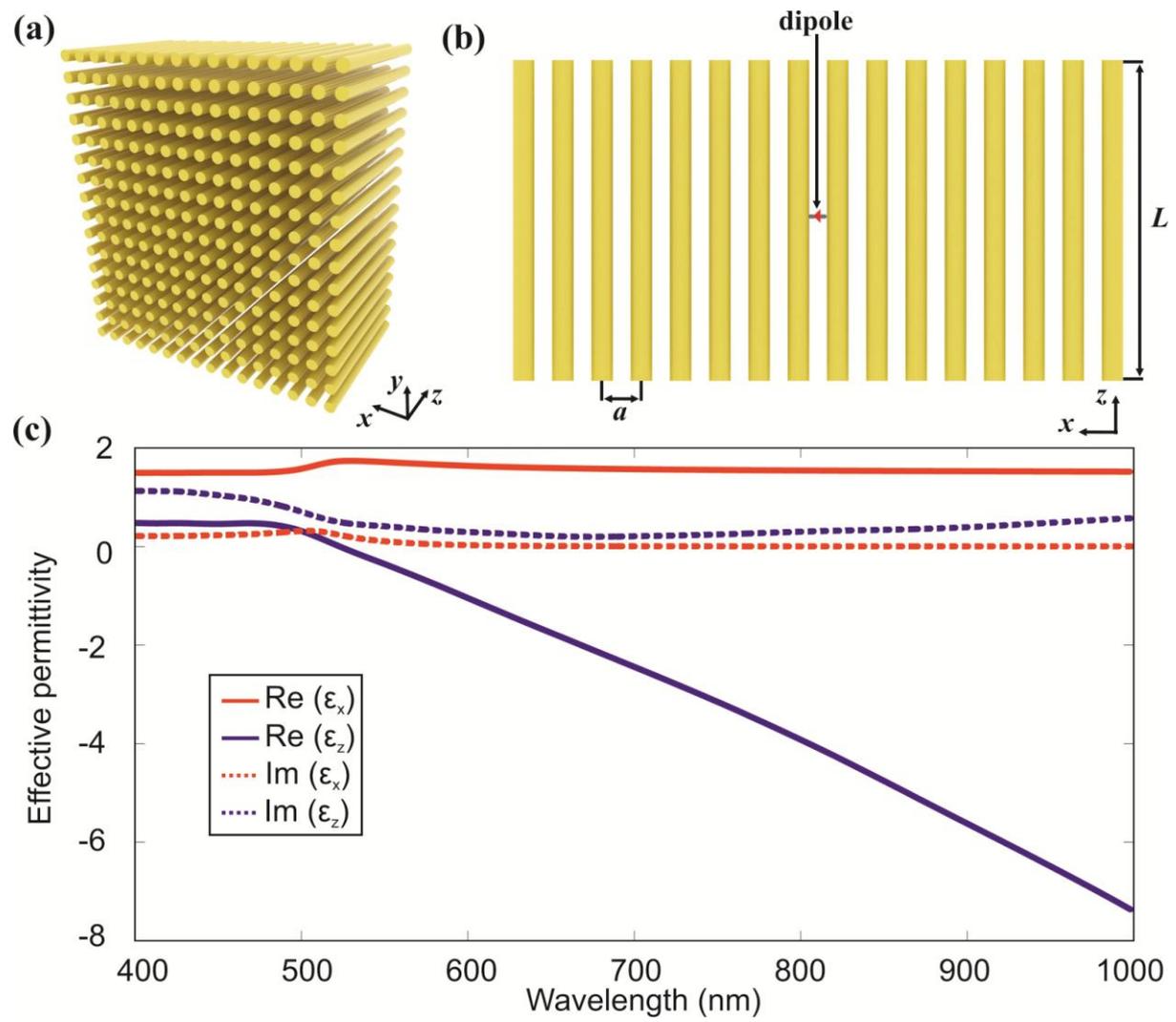

Figure 2

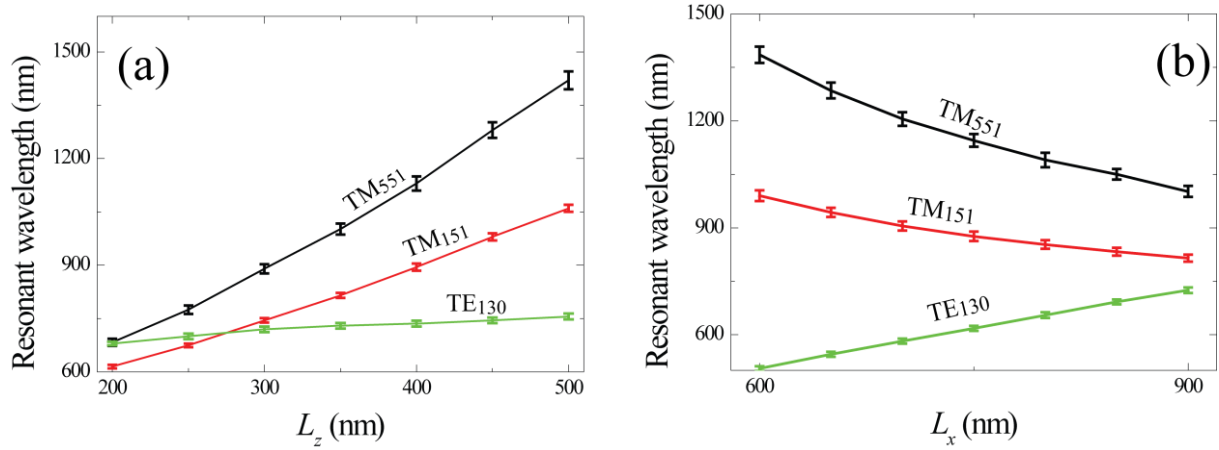

Figure 3

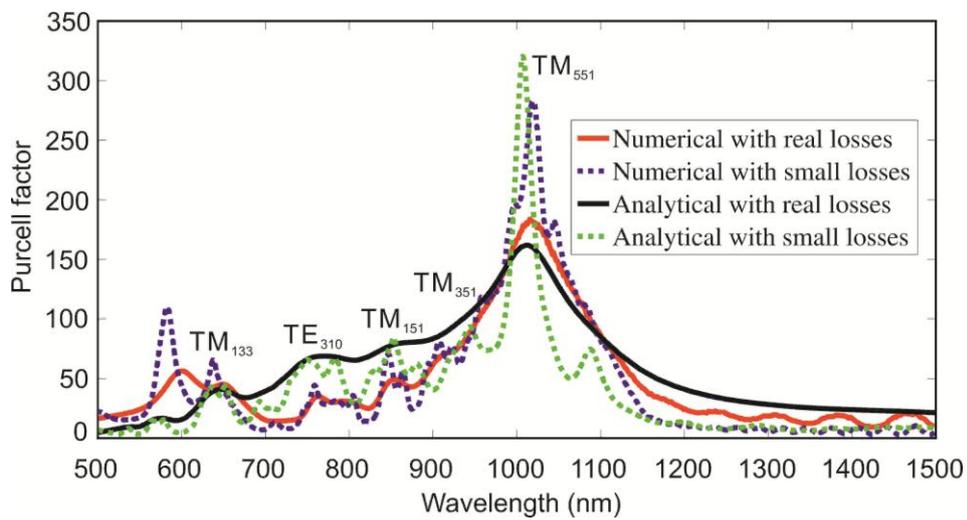

Figure 4

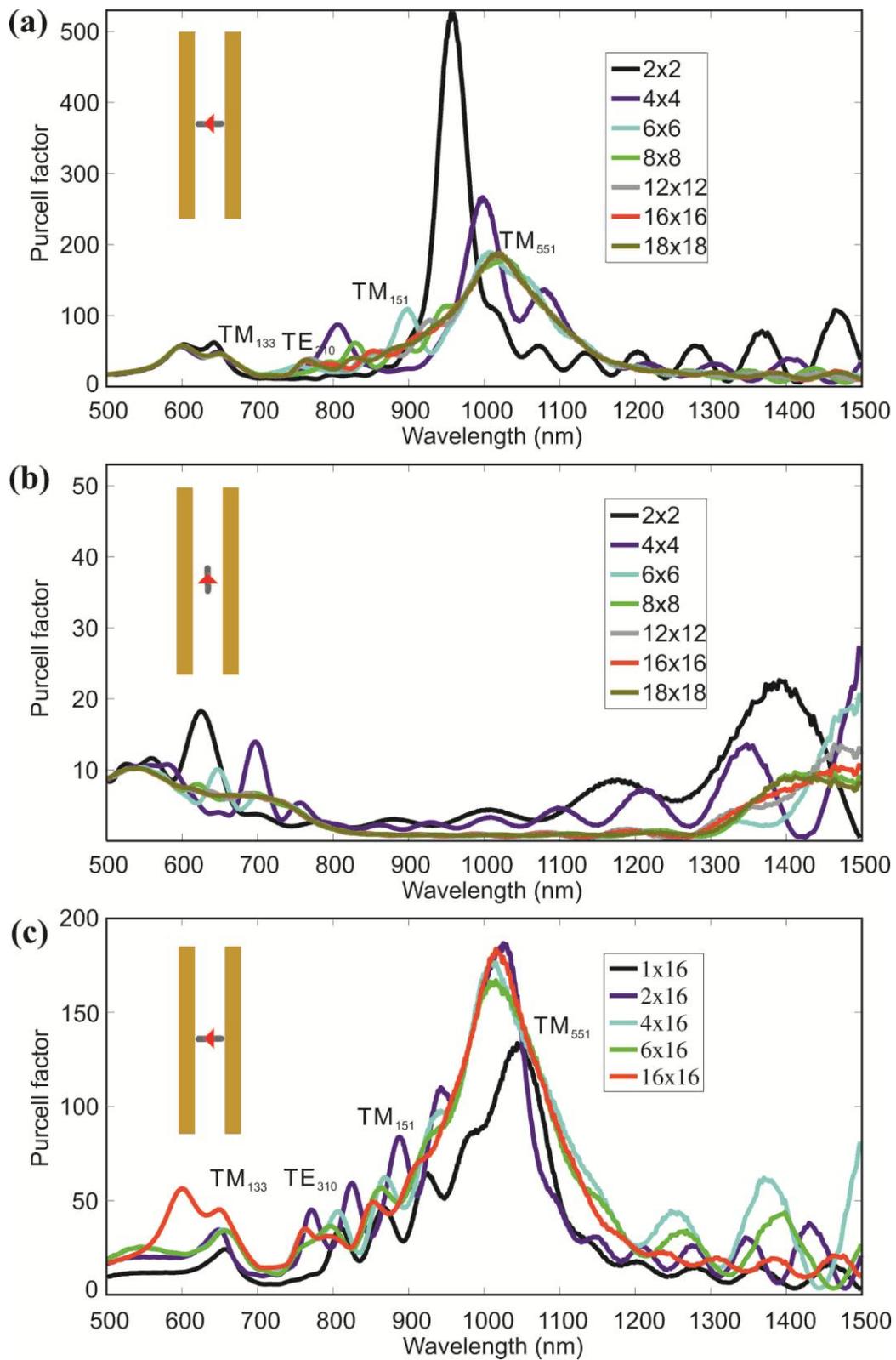

Figure 5

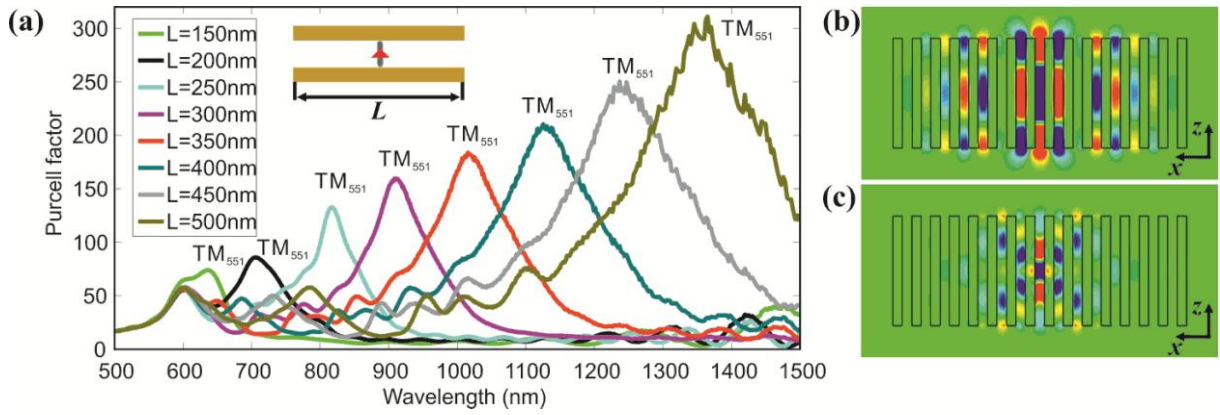

Figure 6

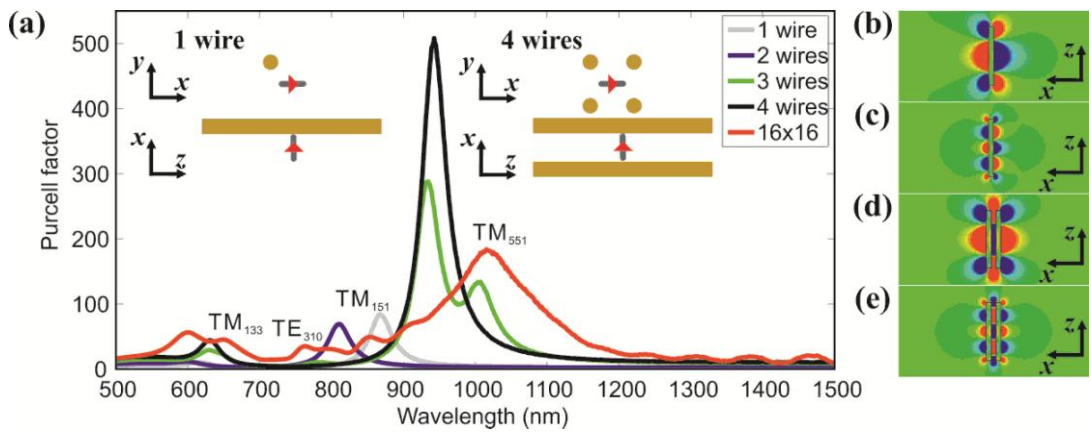

Figure 7.

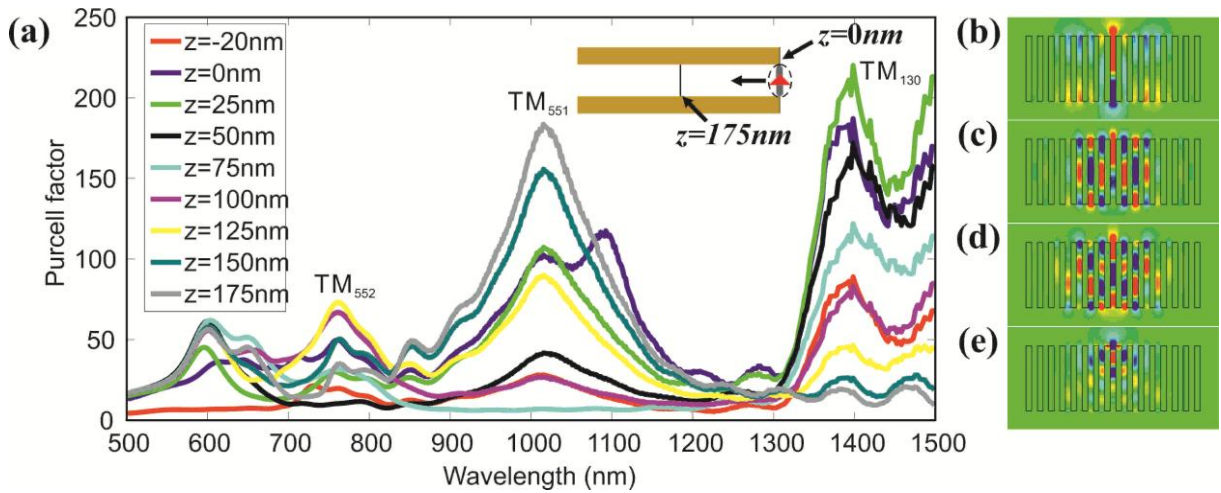

References:

1. L. Novotny and B. Hecht, *Principles of Nano-Optics* (Cambridge University Press 2006).

2. E. M. Purcell, Phys. Rev. **69**, 681 (1946).

3. A. Hayat , P. Ginzburg, and M. Orenstein, Phys. Rev. Lett. **103**, 023601 (2009).

4. A. N. Poddubny, P. Ginzburg, P. A. Belov, A. V. Zayats, and Y. S. Kivshar, Phys. Rev. A **86**, 033826 (2012).

5. K. J. Vahala, Nature **424**, 839-846 (2003).

6. P.Ginzburg, A. V. Krasavin, D. Richards, and A. V. Zayats, Phys. Rev. A **90**, 043836 (2014).

7.D. K. Gramotnev and S.I. Bozhevolnyi, Nature Photon. **4**, 83 - 91 (2010).

8. P.Ginzburg, D.Arbel, and M. Orenstein, Opt. Lett. **31**, 3288-3290 (2006).

9. L. Novotny and N. van Hulst, Nature Photon. **5**, 83–90 (2011).

10. J.-J. Greffet, Science **308**, 1561-1563 (2005).

11. A. V. Akimov, A. Mukherjee, C. L. Yu, D. E. Chang, A. S. Zibrov, P. R. Hemmer, H. Park, and M. D. Lukin, Nature **450**, 402-406 (2007).

12. A.G. Curto, G. Volpe, T.H. Taminiau, M.P. Kreuzer, R. Quidant, N.F. van Hulst, Science **329**, 930 (2010).

13 A.N. Poddubny, P.A. Belov, and Yu. S. Kivshar, "Spontaneous radiation of a finite-size dipole emitter in hyperbolic media", Phys. Rev. A **84**, 023807 (2011).

14. A. Poddubny, I. Iorsh, P. Belov, and Y. Kivshar, Nat. Photonics **7**, 948 (2013).

15. Z. Jacob, J. Kim, G. V. Naik, A. Boltasseva, E. E. Narimanov, and V. M. Shalaev, Appl. Phys. B **100**, 215 (2010).

16. A. N. Poddubny, P. A. Belov, and Y. S. Kivshar, Phys. Rev. B **87**, 035136 (2013).

17. A. N. Poddubny, P. A. Belov, P. Ginzburg, A.V. Zayats, and Y. S. Kivshar, Phys. Rev. B **86**, 035148 (2012).

18. W. Yan, M. Wubs, and N. A. Mortensen, Phys. Rev. B **86**, 205429 (2012).

19. T. Tumkur, G. Zhu, P. Black, Yu. A. Barnakov, C. E. Bonner, and M. A. Noginov, Apll. Phys. Lett. **99**, 151115 (2011).

20. H. N. S.Krishnamoorthy, Z. Jacob, E. Narimanov, I. Kretzschmar, and V. M. Menon, Science **336**, 205–209 (2012).

21.P. Ginzburg, F. J. Rodríguez Fortuño, G. A. Wurtz, W. Dickson, A. Murphy, F. Morgan, R. J. Pollard, I. Iorsh, A. Atrashchenko, P. A. Belov, Y. S. Kivshar, A. Nevet, G. Ankonina, M. Orenstein, and A. V. Zayats, Opt. Express **21**, 14907-14917 (2013).

22. X. Yang, J. Yao, J. Rho, X. Yin, and X. Zhang, Nat. Photonics **6**, 450 (2012).

23. Y. He, S. He and X. Yang, Opt. Lett. **37**, 2907–2909 (2012).

24. B. Wells, A. Zayats, and V. Podolskiy, Phys. Rev. B **89**, 035111 (2014).

25 O. Kidwai, S. V. Zhukovsky, and J. E. Sipe, Opt. Lett. **36**, 2530–2532 (2011).

26. www.cst.com

27. A. P. Slobozhanyuk, A. N. Poddubny, A. E. Krasnok, and P. A. Belov, Appl. Phys. Lett. **104**, 161105 (2014).

28 A. E. Krasnok, A. P. Slobozhanyuk, C. R. Simovski, S. A. Tretyakov, A. N. Poddubny, A. E. Miroshnichenko, Y. S. Kivshar, P. A. Belov, ArXiv e-prints (2015), arXiv:1501.04834 [physics.optics].

29. M. Agio and A. Alu, *Optical Antennas* (Cambridge University Press, 2013).

30. A. K. Okaya and L. F. Barash, Proc. IRE **50**, 2081–2092 (1962).

31. N. Vasilantonakis, M. Nasir, W. Dickson, G. A. Wurtz, A. V. Zayats, Laser Phot. Rev. **9**, DOI: 10.1002/lpor.201400457 (2015).

32. R.K Moniga and A. Ittipiboon, IEEE Transactions on Antennas and Propagation **45**, 1348-1356 (1997).

33 F. Lemoult, M. Fink, G. Lerosey, Nature Communications **3**, 889 (2012).

34 V.Podolskiy, P. Ginzburg, B. Wells and A. Zayats, Faraday Discussions **178**, DOI: 10.1039/C4FD00186A (2014).

35 S.V. Zhukovsky, O. Kidwai, and J. E. Sipe, Optics Express **21**, 14982 (2013).

36.P. B. Johnson, R. W. Christy, "Optical constants of the noble metals," Phys. Rev. B **6**, 4370–4379, (1972).